\documentclass[11pt]{amsart}
\usepackage{amsmath}
\usepackage{amsfonts}
\usepackage{amssymb}
\usepackage{enumerate}
\usepackage{graphicx}
\usepackage{hyperref}

\newtheorem{thm}{Theorem}[section]

\newtheorem{cor}[thm]{Corollary}

\newtheorem{lemma}[thm]{Lemma}
\newtheorem{prop}[thm]{Proposition}

\newtheorem{defn}[thm]{Definition}

\newcommand{\bra}[1]{\langle #1 |}
\newcommand{\ket}[1]{| #1 \rangle}
\newcommand{\braket}[2]{\langle #1 | #2 \rangle}
\newcommand{\ketbra}[2]{| #1 \rangle\langle #2 |}
\newcommand{\Tr}{\mathrm{Tr}}
\newcommand{\bb}[1]{\mathbb{#1}}
\newcommand{\cl}[1]{\mathcal{#1}}

\begin{document}

\title[A Family Of Norms With Applications In Quantum Information Theory II]{A Family Of Norms With Applications In \\ Quantum Information Theory II}

\author[N.~Johnston, D.~W.~Kribs]{Nathaniel Johnston$^1$ and David~W.~Kribs$^{1,2}$}
\address{$^1$Department of Mathematics \& Statistics, University of Guelph,
Guelph, ON, Canada N1G 2W1}
\address{$^2$Institute for Quantum Computing, University of Waterloo, Waterloo, ON, Canada
N2L 3G1}

\begin{abstract}
We consider the problem of computing the family of operator norms recently introduced in \cite{JK09}. We develop a family of semidefinite programs that can be used to exactly compute them in small dimensions and bound them in general. Some theoretical consequences follow from the duality theory of semidefinite programming, including a new constructive proof that there are non-positive partial transpose Werner states that are $r$-undistillable for arbitrary $r$. Several examples are considered via a MATLAB implementation of the semidefinite program, including the case of Werner states and randomly generated states via the Bures measure, and approximate distributions of the norms are provided. We extend these norms to arbitrary convex mapping cones and explore their implications with positive partial transpose states.
\end{abstract}

\maketitle

\section{Introduction}

In \cite{JK09} we initiated the study of a family of operator norms that  quantify the different degrees of entanglement in quantum states. Taken together with previous work \cite{CKo09,CKS09,PPHH07}, where special cases of the norms were considered, these norms have found applications to central problems in quantum information theory. Most importantly, to the problem of determining $k$-entanglement witnesses and $k$-positivity of linear maps, and to the existence problem for non-positive partial transpose (NPPT) bound entangled states. The work \cite{JK09} in particular makes it clear that computing or bounding these norms even in special cases would have a significant impact on these problems.

The primary goal of this paper is to continue the investigation of \cite{JK09}. Here we focus on the development of algorithmic techniques to calculate and bound the operator norms, and we present further applications of the norms in quantum information. Specifically, we develop a family of semidefinite programs \cite{GLS93,Wa09,JJUW09} that can be used to exactly compute them in small dimensions and bound them in general. Some theoretical consequences then follow from the duality theory of semidefinite programming, including a new constructive proof that there are NPPT Werner states \cite{W89} that are $r$-undistillable for arbitrary $r$ \cite{LBCKKSST00}. We consider several examples via a MATLAB implementation of the semidefinite program. In particular, we show how they can be computed on Werner states and randomly generated states via the Bures measure \cite{B69,U76,OSZ09}, and provide approximate distributions of the norms. We also extend these norms to arbitrary convex mapping cones and explore their implications with positive partial transpose states \cite{S86,S08,SSZ09}, and we apply them to a recent conjecture on the regularized relative entropy of entanglement \cite{BRANCONJ,ASW10}.


The paper is arranged as follows. In Section~\ref{sec:prelim} we present our notation and terminology and introduce the reader to the Schmidt rank of vectors and Schmidt number of density operators, $k$-positivity of linear maps, and the Choi-Jamiolkowski isomorphism. The operator norms are defined and some of their most important properties are presented. In Section~\ref{sec:SP} we will present an introduction of semidefinite programs and Section~\ref{sec:semidefProgramMNorm} will follow up by showing how they can be used to compute the operator norms in small dimensions and upper bound them in general. We give a new constructive proof that there are NPPT Werner states that are $r$-undistillable for arbitrary $r$ in Section~\ref{sec:wernerundstil}. Section~\ref{sec:examples} contains MATLAB code that carries out the semidefinite programs, considers examples that demonstrate the performance of the semidefinite programs, and investigates the approximate distribution of the norms in small dimensions. The operator norms are then generalized to a larger family of convex cones in Section~\ref{sec:coneNorms} and their properties are investigated. We conclude in Section~\ref{sec:BrandaoConjecture} with further analysis of how the norms behave in the important case of projections.

\section{Preliminaries}\label{sec:prelim}

Our set-up is similar to that of \cite{JK09} and thus the interested reader is directed there to learn about concepts such as Hermicity-preserving maps, the Choi matrix of a map \cite{C75,Paulsentext}, and the Schmidt number of density operators \cite{TH00}. For us $\cl{H}_n$ denotes an $n$-dimensional complex Hilbert space and $\cl{L}(\cl{H}_n)$ denotes the set of linear operators acting on $\cl{H}_n$. $id_n$ will represent the identity map on $\cl{L}(\cl{H}_n)$. We will consider bipartite systems $\cl{H}_n \otimes \cl{H}_m$ and assume that $m \leq n$. A unit vector $\ket{v} \in \cl{H}$ (a pure state) is denoted using Dirac bra-ket notation, with $\bra{v} := \ket{v}^*$.
We will denote the computational basis vectors (i.e., the vectors with $1$ in the $i^{th}$ component and $0$ in all other components) by $\{\ket{i}\}$.

If $X \in \cl{L}(\cl{H})$ is positive then we will write $X \geq 0$, and $A \geq B$ indicates that $A - B \geq 0$. If it is important to note that a positive operator $X$ is invertible, we will write $X > 0$. It will sometimes be convenient to denote the cone of positive operators in $\cl{L}(\cl{H})$ by $(\cl{L}(\cl{H}))^{+}$. A (mixed) quantum state is represented by a density operator $\rho \geq 0$ that satisfies $\Tr(\rho) = 1$. Whenever lowercase Greek letters like $\rho$ or $\sigma$ are used, it is assumed that they are density operators. General operators will be represented by uppercase letters like $X$ and $Y$. $E := \frac{1}{n}\sum_{ij=0}^{n-1}\ketbra{i}{j}\otimes\ketbra{i}{j}$ refers to the rank-$1$ projection onto the standard maximally entangled state. We will say that a Hermitian operator $X = X^* \in \cl{L}(\cl{H}_n) \otimes \cl{L}(\cl{H}_m)$ is \emph{$k$-block positive} (or equivalently a \emph{$k$-entanglement witness}) if it is the Choi matrix $(id_n \otimes \Phi)(E)$ of a linear map $\Phi : \cl{L}(\cl{H}_n) \rightarrow \cl{L}(\cl{H}_m)$ that is $k$-positive. The scaled Choi matrix of the transpose map, $S := (id_n \otimes T)(nE)$, is a unitary that is referred to as the \emph{swap operator} because $S(\ket{a} \otimes \ket{b}) = \ket{b} \otimes \ket{a}$ for any separable pure state $\ket{a} \otimes \ket{b}$.

\subsection{Relationship Between Schmidt Number and $k$-Positivity}\label{sec:kPosMaps}

Observe that the set of states $\rho$ with $SN(\rho) \leq k$ is a closed convex cone if we remove the requirement that $\Tr(\rho) = 1$. This cone will be denoted $\cl{S}_k$. Given a convex cone $\cl{C} \subseteq \cl{L}(\cl{H})$, its \emph{dual cone} is the convex cone defined through the Hilbert-Schmidt inner product as follows:
\[
  \cl{C}^O := \big\{ X \in \cl{L}(\cl{H}) : \Tr(XY) \geq 0 \quad \forall \, Y \in \cl{C} \big\}.
\]

\noindent It is known that the dual cone of $\cl{S}_k$, the operators with Schmidt number no greater than $k$, is exactly the set of $k$-block positive operators, and vice-versa \cite{S08,SSZ09}.

The following two theorems are each just a way of restating the fact that the cone of unnormalized states with Schmidt number at most $k$ is dual to the cone of $k$-block positive operators. Theorem~\ref{thm:kPos} in particular provides an important second characterization of $k$-positivity of a linear map that is sometimes given as the definition of $k$-positivity in quantum information theory~\cite{TH00}.
\begin{thm}\label{thm:kPos}
	If $\Phi : \cl{L}(\cl{H}_n) \rightarrow \cl{L}(\cl{H}_n)$ is a linear map, then $\Phi$ is $k$-positive if and only if
	\begin{align*}
		(id_n \otimes \Phi)(\rho) \geq 0 \quad \forall \, \rho \in \cl{L}(\cl{H}_n) \otimes \cl{L}(\cl{H}_n) \text{ with } SN(\rho) \leq k.
	\end{align*}
\end{thm}

\begin{thm}\label{thm:thmain}
	If $\rho \in \cl{L}(\cl{H}_n) \otimes \cl{L}(\cl{H}_n)$ is a density operator, then $SN(\rho) \leq k$ if and only if
	\begin{align*}
		(id_n \otimes \Phi)(\rho) \geq 0 \quad \forall \, k\text{-positive } \Phi : \cl{L}(\cl{H}_n) \rightarrow \cl{L}(\cl{H}_n).
	\end{align*}
\end{thm}

These theorems are of theoretical interest, but are not of much practical use for testing $k$-positivity or Schmidt number, since (for example) it is not possible to apply $id_n \otimes \Phi$ to $\rho$ and check positivity for every $k$-positive map $\Phi$. In some cases, however, an explicit finite set $\cl{S}$ of $k$-positive maps is known for which $(id_n \otimes \Phi)(\rho) \geq 0$ for all $\Phi \in \cl{S}$ implies $SR(\rho) \leq k$. For example, if $m = 2$ and $n = 2$ or $n = 3$ then the transpose map $T$ alone is enough to determine whether or not $\rho$ is separable (i.e., $SN(\rho) = 1$) \cite{HHH96}. The fact that the transpose map can be used to determine separability in small dimensions has led to the study of \emph{positive partial transpose (PPT)} states \cite{P96}, which are density operators $\rho$ such that $(id_n \otimes T)(\rho) \geq 0$. Throughout the rest of this paper, we will write the partial transpose operation $(id_n \otimes T)(\rho)$ as $\rho^\Gamma$.

The connections between $k$-positivity and Schmidt number via dual cones have been studied quite a bit over the last few years. This basic theme of duality between $k$-positivity and Schmidt number will be present throughout much of this paper.

\subsection{Family of Operator Norms}\label{sec:OperatorNorms}

For $1 \leq k \leq m$, the $k$th operator norm introduced in \cite{JK09} has the following form for positive operators $X$:
\begin{align*}
 	\big\|X\big\|_{S(k)} & = \sup_{\ket{v}} \Big\{ \bra{v}X\ket{v} : SR(\ket{v}) \leq k \Big\} \\
 	 & = \sup_{\rho} \Big\{ \Tr(X\rho) : SN(\rho) \leq k \Big\}.
\end{align*}

The following two simple results about these norms were proved in \cite{JK09}. Proposition~\ref{prop:kPosInf1} shows that the problem of computing the norms for positive operators is equivalent to the problem determining $k$-block positivity for arbitrary operators and is thus likely very difficult. Proposition~\ref{prop:rankOneNorm} shows that we can nonetheless efficiently compute the operator norms when the operator under consideration has rank $1$.

\begin{prop}\label{prop:kPosInf1}
  Let $X \in (\cl{L}(\cl{H}_n) \otimes \cl{L}(\cl{H}_m))^{+}$ be positive and let $c \in \bb{R}$. Then $cI - X$ is $k$-block positive if and only if $c \geq \big\|X\big\|_{S(k)}$.
\end{prop}

\begin{prop}\label{prop:rankOneNorm}
	Let $\rho = \ketbra{v}{v} \in \cl{L}(\cl{H}_n) \otimes \cl{L}(\cl{H}_m)$ be a pure state and let $\{ \alpha_i \}$ be the Schmidt coefficients of $\ket{v}$ ordered so that $\alpha_1 \geq \alpha_2 \geq \cdots \geq \alpha_m \geq 0$. Then
	\begin{align*}
	  \big\| \rho \big\|_{S(k)} & = \sum_{i=1}^k \alpha_i^2.
  \end{align*}
\end{prop}

\subsection{Semidefinite Programming}\label{sec:SP}

Here we introduce the reader to semidefinite programming (SP), which we will see provides a step in the direction of being able to compute the operator norms defined above. Our introduction will be brief -- for a more in-depth introduction and discussion, the reader is encouraged to read any of a number of other sources including \cite{Al95,VB96,Lo03,Kl02,WSV00}. Most importantly, there are explicit methods that are able to approximately solve semidefinite programs of the type presented in this paper to any desired accuracy in polynomial time \cite{GLS93}.

For our purposes, assume we have a Hermicity-preserving linear map $\Phi : \cl{L}(\cl{H}_n) \rightarrow \cl{L}(\cl{H}_m)$, two operators $A \in \cl{L}(\cl{H}_n)$ and $B \in \cl{L}(\cl{H}_m)$, and a convex cone $\cl{C} \subseteq (\cl{L}(\cl{H}_n))^+$. Then the corresponding semidefinite program is given by the following pair of optimization problems:
\begin{align}\label{sp:form}
\begin{matrix}
\begin{tabular}{r l c r l}
\multicolumn{2}{c}{{\bf Primal problem}} & \quad \quad \quad & \multicolumn{2}{c}{{\bf Dual problem}} \\
\text{maximize:} & $\Tr(AX)$ & \quad & \text{minimize:} & $\Tr(BY)$ \\
\text{subject to:} & $\Phi(X) \leq B$ & \quad & \text{subject to:} & $\Phi^\dagger(Y) \geq A$ \\
\ & $X \in \cl{C}$ & \ & \ & $Y \in \cl{C}^O$ \\
\end{tabular}
\end{matrix}
\end{align}

Though the semidefinite program~\eqref{sp:form} differs from the standard form of semidefinite programs, it is equivalent and better suited to our particular needs. This form has been used very recently to solve other problems in quantum information \cite{Wa09,JJUW09}. The interested reader is pointed to Appendix I for a discussion of how to convert between the form \eqref{sp:form} and the standard form, as well as MATLAB code that performs the conversion in order to allow pre-existing software to solve these semidefinite programs of the form~\eqref{sp:form}.

We define the \emph{primal feasible} set $\cl{A}$ and \emph{dual feasible} set $\cl{B}$ to be
\begin{align*}
	\cl{A} := \big\{ X \in \cl{C} : \Phi(X) \leq B \big\} \quad \quad \text{ and } \quad \quad \cl{B} := \big\{ Y \in \cl{C}^O : \Phi^\dagger(Y) \geq A \big\}.
\end{align*}

\noindent The optimal values associated with the primal and dual problems are defined to be
\begin{align*}
	\alpha := \sup_{X \in \cl{A}} \big\{ \Tr(AX) \big\} \quad \quad \text{ and } \quad \quad \beta := \inf_{Y \in \cl{B}} \big\{ \Tr(BX) \big\},
\end{align*}

\noindent and if $\cl{A}$ or $\cl{B}$ is empty then we set $\alpha = -\infty$ or $\beta = \infty$, respectively.

Semidefinite programming has a strong theory of duality. The theory of weak duality tells us it is always the case that $\alpha \leq \beta$. Equality is actually attained for many semidefinite programs of interest though, as the following theorem shows.
\begin{thm}[Strong duality]\label{thm:spSlater}
  The following two implications hold for every semidefinite program of the form~\eqref{sp:form}.
  \begin{enumerate}[1.]
		\item Strict primal feasibility: If $\beta$ is finite and there exists an operator $X$ in the interior of $\cl{C}$ such that $\Phi(X) < B$, then $\alpha = \beta$ and there exists $Y \in \cl{B}$ such that $\Tr(YB) = \beta$.
    \item Strict dual feasibility: If $\alpha$ is finite and there exists an operator $Y$ in the interior of $\cl{C}^O$ such that $\Phi^\dagger(Y) > A$, then $\alpha = \beta$ and there exists $X \in \cl{A}$ such that $\Tr(XA) = \alpha$.
  \end{enumerate}
\end{thm}

There are other conditions that imply strong duality, but the conditions of Theorem~\ref{thm:spSlater} (which are known as \emph{Slater-type conditions}) will be sufficient for our needs.

\section{Bounding The Operator Norms}\label{sec:semidefProgramMNorm}

Proposition~\ref{prop:rankOneNorm} shows that we can compute the $k$th operator norms of rank-$1$ operators efficiently, since the Schmidt coefficients of a vector can be computed in $O(n^3)$ time. However, Proposition~\ref{prop:kPosInf1} shows that the problem of computing these  operator norms for arbitrary positive operators is equivalent to the problem of determining $k$-block positivity for arbitrary operators and is thus likely very difficult. Here we develop a family of semidefinite programs that can be used to provide upper bounds on the norms in general and compute them exactly in low-dimensional cases. Additionally, some simple theoretical results that further establish the link between the $k$th operator norm and $k$-block positive operators will follow from the duality theory of semidefinite programming.

Given a positive operator $X \in (\cl{L}(\cl{H}_n) \otimes \cl{L}(\cl{H}_m))^+$ and a natural number $k$, we now present a family of semidefinite programs with the following properties:
\begin{itemize}
	\item Strong duality holds for each semidefinite program.
	\item The optimal value $\alpha$ of each SP is an upper bound of $\big\|X\big\|_{S(k)}$.
	\item There is an SP in the family such that the optimal value satisfies $\alpha = \big\|X\big\|_{S(k)}$.
\end{itemize}

Let $X \in (\cl{L}(\cl{H}_n) \otimes \cl{L}(\cl{H}_m))^+$ be a positive operator for which we wish to compute $\big\|X\big\|_{S(k)}$. Let $\Phi_k : \cl{L}(\cl{H}_m) \rightarrow \cl{L}(\cl{H}_m)$ be a fixed $k$-positive linear map and consider the following semidefinite program:
\begin{align}\label{sp:matrixNorm}
\begin{matrix}
\begin{tabular}{r l c r l}
\multicolumn{2}{c}{{\bf Primal problem}} & \quad \quad \quad & \multicolumn{2}{c}{{\bf Dual problem}} \\
\text{maximize:} & $\Tr(X\rho)$ & \quad & \text{minimize:} & $\lambda$ \\
\text{subject to:} & $(id_n \otimes \Phi_k)(\rho) \geq 0$ & \quad & \text{subject to:} & $\lambda I_n \otimes I_m \geq (id_n \otimes \Phi_k^\dagger)(Y) + X$ \\
\ & $\Tr(\rho) = 1$ & \ & \ & $Y \geq 0$ \\
\ & $\rho \geq 0$ & \ & \ & \ \\
\end{tabular}
\end{matrix}
\end{align}

It may not be immediately obvious that this semidefinite program is in the form of \eqref{sp:form}, so we first check that these problems are indeed duals of each other and form a valid semidefinite program. To this end, consider the linear map $\Psi : \cl{L}(\cl{H}_n) \otimes \cl{L}(\cl{H}_m) \rightarrow (\cl{L}(\cl{H}_n) \otimes \cl{L}(\cl{H}_m)) \oplus \cl{L}(\cl{H}_1)$ defined by
\begin{align*}
  \Psi(\rho) = \begin{bmatrix}-(id_n \otimes \Phi_k)(\rho) & 0 \\ 0 & \Tr(\rho) \end{bmatrix}.
\end{align*}

\noindent Then the dual map $\Psi^\dagger : (\cl{L}(\cl{H}_n) \otimes \cl{L}(\cl{H}_m)) \oplus \cl{L}(\cl{H}_1) \rightarrow \cl{L}(\cl{H}_n) \otimes \cl{L}(\cl{H}_m)$ is given by
\begin{align*}
  \Psi^\dagger\Big( \begin{bmatrix} Y & * \\ * & \lambda \end{bmatrix} \Big) = \lambda I_n \otimes I_m - (id_n \otimes \Phi_k^\dagger)(Y).
\end{align*}

\noindent Finally, setting
\[
  A := X \quad \text{and} \quad B := \begin{bmatrix} 0 & 0 \\ 0 & 1 \end{bmatrix}
\]

\noindent and recalling that the convex cone of positive semidefinite operators is its own dual cone gives the semidefinite program~\eqref{sp:matrixNorm} in standard form.

We now show that this program satisfies the Slater-type conditions for strong duality given by Theorem~\ref{thm:spSlater}. It is clear that both $\alpha$ and $\beta$ are finite, as $\Tr(X\rho) \leq \big\| X \big\|$ and $\lambda \geq 0$. Both feasible sets are also non-empty (for example, one could take $\rho$ to be any separable state, $Y = 0$, and $\lambda \geq \big\| X \big\|$). Strong dual feasibility then follows by choosing any $Y > 0$ and a sufficiently large $\lambda$. Strong primal feasibility is not necessarily satisfied, however, as there is no guarantee that $\Phi_k$ does not introduce singularities in $\rho$ (for example, consider the zero map, which is $k$-positive). We could restrict the family of $k$-positive maps that we are interested in if we really desired strong primal feasibility, but strict dual feasibility is enough for our purposes.

It follows from Theorem~\ref{thm:kPos} that, for any $k$-positive map $\Phi_k$, the optimal value of the semidefinite program~\eqref{sp:matrixNorm} is an upper bound of $\big\| X \big\|_{S(k)}$ -- the supremum in the primal problem is just being taken over a set that is larger than the set of operators $\rho$ with $SN(\rho) \leq k$. This leads to the following theorem.

\begin{thm}\label{thm:kPosInf2}
	Let $X \in (\cl{L}(\cl{H}_n) \otimes \cl{L}(\cl{H}_m))^{+}$. Then
	\begin{align*}
	  \big\| X \big\|_{S(k)} = \inf_{Y} \big\{ \big\|X + Y\big\| : Y\text{ is }k\text{-block positive} \big\}.
	\end{align*}
\end{thm}
\begin{proof}
	Because $\Phi_k$ is $k$-positive if and only if $\Phi_k^\dagger$ is $k$-positive, the dual problem~\eqref{sp:matrixNorm} can be rephrased as asking for the infimum of $\big\| X + Y \big\|$, where the infimum is taken over a subset of the $k$-block positive operators $Y \in \cl{L}(\cl{H}_n) \otimes \cl{L}(\cl{H}_m)$. The preceding paragraph then showed us that
	\[
		\big\| X \big\|_{S(k)} \leq \inf_{Y} \big\{ \big\|X + Y\big\| : Y\text{ is }k\text{-block positive} \big\}.
	\]
	
	\noindent To see that equality is attained, choose $Y := \big\| X \big\|_{S(k)}I - X$, which we know from Proposition~\ref{prop:kPosInf1} is $k$-block positive. Then
	\[
		\big\|X + Y\big\| = \big\|X + \big\| X \big\|_{S(k)}I - X\big\| = \big\| X \big\|_{S(k)}.
	\]
\end{proof}

In fact, it is not difficult to see that there is a particular $k$-positive map $\Phi_k$ such that $\big\| X \big\|_{S(k)}$ is attained as the optimal value of the semidefinite program~\eqref{sp:matrixNorm} corresponding to $\Phi_k$ -- simply let $\Phi_k$ be the map associated with the operator $\big\| X \big\|_{S(k)}I - X$ via the Choi-Jamiolkowski isomorphism.

One additional obvious implication of Theorem~\ref{thm:kPosInf2} is that $\big\| X \big\|_{S(k)} \leq \big\|X + Y\big\|$ for all $X \in (\cl{L}(\cl{H}_n) \otimes \cl{L}(\cl{H}_m))^+$ and all $k$-block positive $Y \in \cl{L}(\cl{H}_n) \otimes \cl{L}(\cl{H}_m)$. The following corollary shows that this can be strengthened into another characterization of $k$-positivity.

\begin{cor}\label{cor:SPcor}
	Let $Y = Y^* \in \cl{L}(\cl{H}_n) \otimes \cl{L}(\cl{H}_m)$. Then $Y$ is $k$-block positive if and only if
	\begin{align*}
	  \big\| X \big\|_{S(k)} \leq \big\|X + Y\big\| \quad \forall \, X \in (\cl{L}(\cl{H}_n) \otimes \cl{L}(\cl{H}_m))^{+}.
	\end{align*}
\end{cor}
\begin{proof}
	The ``only if'' direction of the proof follows immediately from Theorem~\ref{thm:kPosInf2}. To see the ``if'' direction, assume that $Y$ is not $k$-block positive and choose $X = cI - Y$, where $c \in \bb{R}$ is large enough that $cI - Y \geq 0$. Then, because $Y$ is not $k$-block positive, there exists a vector $\ket{v}$ with $SR(\ket{v}) \leq k$ such that $\bra{v}Y\ket{v} < 0$. Thus
	\[
		\big\| X \big\|_{S(k)} \geq \bra{v}(cI - Y)\ket{v} = c - \bra{v}Y\ket{v} > c = \big\|X + Y\big\|.
	\]
\end{proof}

Recall that if $m = 2$ and $n = 2$ or $n = 3$ then the transpose map $T$ alone is enough to determine whether or not $\rho$ is separable (i.e., $SN(\rho) = 1$ if and only if $(id_n \otimes T)(\rho) \geq 0$). It follows that the semidefinite program~\eqref{sp:matrixNorm} with $k = 1$ and $\Phi_1 = T$ can be used to compute $\big\| X \big\|_{S(1)}$ for positive operators $X$. That is, the infinite family of semidefinite programs reduces to just a single semidefinite program in this situation. We can then use Proposition~\ref{prop:kPosInf1} to determine $1$-block positivity of operators $X \in \cl{L}(\cl{H}_3) \otimes \cl{L}(\cl{H}_2)$.

\section{Undistillable Werner States}\label{sec:wernerundstil}

Given a bipartite state $\rho$, a natural question to ask is whether of not it can be transformed via local operations and classical communication to a maximally entangled pure state. If it can, $\rho$ is said to be $1$-distillable. It may happen that $\rho$ itself cannot be transformed into a maximally entangled pure state in this way, but $r$ copies of $\rho$ can. In this situation, $\rho$ is said to be \emph{$r$-distillable}. If there exists an $r$ such that $\rho$ is $r$-distillable, then $\rho$ is simply called \emph{distillable} and otherwise $\rho$ is said to be {\it undistillable}. The key connection between (un)distillability and the family of norms considered here comes from a result of \cite{DSSTT00,DCL00} that says a state $\rho$ is $r$-undistillable if and only if $(\rho^{\otimes r})^\Gamma$ is $2$-block positive. We will thus focus particularly on the $2$-norm in this section.

Not surprisingly, separable states are undistillable. It is also known that all states with positive partial transpose are undistillable \cite{H97,HHH98}, but the converse remains an open problem. That is, are there states with non-positive partial transpose (NPPT) that are undistillable? In this section we compute the $1$-norm on a family of projections, and use that result to find states that are $r$-undistillable for arbitrarily large $r$. Note that this does not answer the question on whether or not NPPT undistillable states exist, however, because in our construction, the viable values of $r$ are bounded above by a function of the dimension ($n$) of the Hilbert space.

We shall focus on Werner states \cite{W89}, which are a particular family of bipartite states in $\cl{L}(\cl{H}_n) \otimes \cl{L}(\cl{H}_n)$ that are central in quantum information. They are exactly the states that are invariant under any operator of the form $U \otimes U$, where $U \in \cl{L}(\cl{H}_n)$ is unitary, and they take the following form:
	\[
		\rho_\alpha := \frac{1}{n(n - \alpha)}(I - \alpha S) \in \cl{L}(\cl{H}_n) \otimes \cl{L}(\cl{H}_n) \ \text{ for some } \alpha \in [-1,1],
	\]
\noindent where $S$ is the swap operator as defined earlier. Werner states have become a subject of great interest because it has been shown that NPPT undistillable states exist if and only if there is an NPPT undistillable Werner state \cite{HH98}. The following is a well-known result on Werner states.

\begin{lemma}\label{prop:rUndistMonotone}
	If $\rho_{\alpha^\prime}$ is $r$-undistillable, then $\rho_\alpha$ is $r$-undistillable for all $\alpha \leq \alpha^\prime$.
\end{lemma}

In \cite{JK09} it was shown that the Werner state $\rho_{2/n}$ (which has non-positive partial transpose) is $r$-undistillable if and only if the $2$-norm of the projection onto the negative eigenspace of $(\rho^{\otimes r})^\Gamma$ is less than or equal to $1/2$. More specifically, the projections of interest can be defined recursively as follows:
\begin{align}\label{eq:proj_def}\begin{split}
	{}_nP_1 & := E \in \cl{L}(\cl{H}_n) \otimes \cl{L}(\cl{H}_n), \\
	{}_nP_r & := (I - E) \otimes {}_{n}P_{r-1} + E \otimes (I - {}_{n}P_{r-1}) \quad \forall \, r \geq 2,
\end{split}\end{align}
\noindent where $I$ is the identity operator of appropriate size. Although the $2$-norm of these projections is still unknown, we can compute the $1$-norm of each of these projections using the semidefinite program~\eqref{sp:matrixNorm}.




\begin{lemma}\label{lem:S1}
	Let ${}_nP_r$ be a projection as defined by the recurrence relations~\eqref{eq:proj_def}. Then
	\begin{align*}
		\big\|{}_nP_r\big\|_{S(1)} = \frac{1}{2} - \frac{1}{2}\Big(1 - \frac{2}{n}\Big)^r.
	\end{align*}
\end{lemma}

\begin{proof}
	To see the ``$\geq$'' inequality, consider the separable vector $\ket{v} := \ket{0} \otimes \ket{0} \in \cl{H}_n^{\otimes 2} \otimes \cl{H}_n^{\otimes {2r-2}}$. Then define the quantity
	\begin{align*}
		c_{n,r} := \bra{v} {}_nP_r \ket{v}.
	\end{align*}
It follows that
	\begin{align*}
		c_{n,r} & = \bra{0}(I - E)\ket{0} \bra{0}{}_{n}P_{r-1}\ket{0} + \bra{0}E\ket{0} \bra{0}(I - {}_{n}P_{r-1})\ket{0} \\
		 & = \frac{n-1}{n}c_{n,r-1} + \frac{1}{n}(1 - c_{n,r-1}) \\
		 & = \Big(1 - \frac{2}{n}\Big)c_{n,r-1} + \frac{1}{n}.
	\end{align*}
Standard methods for solving recurrence relations yields $c_{n,r} = \frac{1}{2} - \frac{1}{2}\Big(1 - \frac{2}{n}\Big)^r$. Noting that $\big\|{}_nP_r\big\|_{S(1)} \geq c_{n,r}$ gives the desired inequality.
	
	To see the ``$\leq$'' inequality, we will use the dual form of the semidefinite program~\eqref{sp:matrixNorm} with the transpose map $\Phi_1(X) := X^T$. To this end, notice that if $\lambda_{n,r}^{max}$ is the maximal eigenvalue of ${}_nP_r^\Gamma$, then $(\lambda_{n,r}^{max}I - {}_nP_r^\Gamma)$ is positive semidefinite and so
Theorem~\ref{thm:kPosInf2} says that
	\begin{align*}
		\big\|{}_nP_r\big\|_{S(1)} \leq \big\| {}_nP_r + (\lambda_{n,r}^{max}I - {}_nP_r^\Gamma)^\Gamma   \big\| = \big\| \lambda_{n,r}^{max}I \big\| = \lambda_{n,r}^{max}.
	\end{align*}
In order to compute $\lambda_{n,r}^{max}$, let us consider the partial transpose of the family of projections~\eqref{eq:proj_def}:
\begin{align*}
	{}_nP_1^\Gamma & = \frac{1}{n}S \in \cl{L}(\cl{H}_n) \otimes \cl{L}(\cl{H}_n), \\
	{}_nP_r^\Gamma & = \frac{1}{n}S \otimes (I - {}_{n}P_{r-1}^\Gamma) + (I - \frac{1}{n} S) \otimes {}_{n}P_{r-1}^\Gamma \quad \forall \, r \geq 2.
\end{align*}
It is clear that the eigenvectors of ${}_nP_r^\Gamma$ are each of the form $\ket{x} \otimes \ket{y}$ for some eigenvector $\ket{x}$ of $S$ and some eigenvector $\ket{y}$ of ${}_nP_{r-1}^\Gamma$. If we recall that the eigenvalues of $S$ are $\pm 1$, it follows that
	\begin{align*}
		\lambda_{n,r}^{max} = \max\Big\{ (1 - \frac{2}{n})\lambda_{n,r-1}^{max} + \frac{1}{n} , (1 + \frac{2}{n})\lambda_{n,r-1}^{max} - \frac{1}{n} \Big\}.
	\end{align*}
If $\lambda_{n,r-1}^{max} \leq \frac{1}{2}$ then $(1 + \frac{2}{n})\lambda_{n,r-1}^{max} - \frac{1}{n} \leq (1 - \frac{2}{n})\lambda_{n,r-1}^{max} + \frac{1}{n} \leq \frac{1}{2}$, so it follows via induction (and the fact that $\lambda_{n,1}^{max} = \frac{1}{n} \leq \frac{1}{2}$) that $\lambda_{n,r}^{max} = (1 - \frac{2}{n})\lambda_{n,r-1}^{max} + \frac{1}{n}$. We already saw that this recurrence relation has the closed form $\lambda_{n,r}^{max} = \frac{1}{2} - \frac{1}{2}\Big(1 - \frac{2}{n}\Big)^r$, which finishes the proof.
\end{proof}

We can now state and prove the main result of this section. We use the above results together with results from \cite{JK09} in the proof.

\begin{thm}\label{thm:main}
Let $n,r \in \mathbb{N}$ be such that $p := \frac{(n-2)^r}{n^r - (n-2)^r} \geq 1$. If $r$ is odd and $\alpha \leq \frac{1}{n}(\sqrt[r]{p} + 1)$ or if $r$ is even and $\alpha \leq \frac{1}{n}(\sqrt[r-1]{p} + 1)$, then the Werner state $\rho_\alpha \in \cl{L}(\cl{H}_n) \otimes \cl{L}(\cl{H}_n)$ is $r$-undistillable.
\end{thm}

\begin{proof}
As $\| {}_nP_r \|_{S(2)} \leq 2 \| {}_nP_r \|_{S(1)} $ (see Theorem~4.13 of \cite{JK09}), Lemma~\ref{lem:S1} above implies that
\begin{align}\label{eq:S2ineq}
	\| {}_nP_r \|_{S(2)} \leq 1 - \left(1 - \frac{2}{n}\right)^r.
\end{align}

\noindent The eigenvalues of $(\rho_\alpha^{\otimes r})^\Gamma$ are
\begin{align*}
	(1 - \alpha n)^m	\quad \text{for $m = 0, 1, \ldots, r$}.
\end{align*}
In particular, $(\rho_\alpha^{\otimes r})^\Gamma$ is nonsingular and has some positive eigenvalues, so the first half of condition (2) (with $k=1$) of Theorem~5.1 from \cite{JK09} is satisfied. Now assume that $p := \frac{(n-2)^r}{n^r - (n-2)^r} \geq 1$ and that $\alpha = \frac{1}{n}(\sqrt[2\lceil r/2\rceil - 1]{p} + 1)$. If we can show that $\rho_\alpha$ is $r$-undistillable then we are done by Lemma~\ref{prop:rUndistMonotone}. Well, $p \geq 1$ implies that $\alpha \geq \frac{2}{n}$, so the minimal positive eigenvalue $\lambda_{min}^{+}$ of $(\rho_\alpha^{\otimes r})^\Gamma$ is $1$, and its maximal (in absolute value) negative eigenvalue $\lambda_{max}^{-}$ is $(1 - \alpha n)^{2\lceil r/2\rceil - 1}$. We have
\begin{align*}
	\alpha & = \frac{1}{n}(\sqrt[2\lceil r/2\rceil - 1]{p} + 1) = \frac{1}{n}\Big(\sqrt[2\lceil r/2\rceil - 1]{\frac{(n-2)^r}{n^r - (n-2)^r}} + 1\Big).
\end{align*}

\noindent Rearranging this expression yields, in the second equality,
\begin{align*}
	\lambda_{min}^{+} = 1 & = (\alpha n - 1)^{2\lceil r/2\rceil - 1}\left(\Big(\frac{n}{n-2}\Big)^r - 1\right) \geq \lambda_{max}^{-}\frac{\| {}_nP_r \|_{S(2)}}{1 - \| {}_nP_r \|_{S(2)}}.
\end{align*}

\noindent and where the final inequality comes from~\eqref{eq:S2ineq}. Now by condition (2) of Theorem 5.1 from \cite{JK09}, we have that $(\rho_\alpha^{\otimes r})^\Gamma$ is $2$-block positive, and hence the result follows.
\end{proof}

Note that the value $p$ of Theorem~\ref{thm:main} is such that $p \geq 1$ if and only if $n \geq \frac{2\sqrt[r]{2}}{\sqrt[r]{2}-1}$. Thus, for any $r \geq 1$, there is always some non-PPT Werner state that is $r$-undistillable as long as the dimension $n$ is large enough. In fact, the dimension grows roughly linearly: $\frac{2\sqrt[r]{2}}{\sqrt[r]{2}-1}$ is asymptotic to $\frac{2}{\ln(2)}r + 1$. Also, if $p \geq 1$ then the result immediately implies that the $\alpha = 2/n$ Werner state is $r$-undistillable. Additionally, it is not difficult to see that if $\rho_{\alpha} \in \cl{L}(\cl{H}_n) \otimes \cl{L}(\cl{H}_n)$ is $r$-undistillable then $\rho_\alpha \in \cl{L}(\cl{H}_m) \otimes \cl{L}(\cl{H}_m)$ must also be $r$-undistillable for any $m \leq n$. Putting these facts together gives us the following slightly weaker (but much simpler) corollary of Theorem~\ref{thm:main}.

%
\begin{cor}
	If $\alpha \leq \min\{2/n, \ln(2)/(r + 3\ln(2) - 1)\}$ then $\rho_{\alpha} \in \cl{L}(\cl{H}_n) \otimes \cl{L}(\cl{H}_n)$ is $r$-undistillable.
\end{cor}

	Similar results about $r$-undistillability of Werner states have appeared in the literature in the past. In \cite{LBCKKSST00} it was shown that, for any fixed $n \geq 3$, there exist NPPT Werner states that are $r$-undistillable, though the region that was shown to be $r$-undistillable shrinks exponentially with $r$. Our result is stronger in that our regions shown to be $r$-undistillable shrink only like $1/r$. On the other hand, for each fixed $n$ our result only gives a region of NPPT $r$-undistillability for $r \leq \ln(2)(n - 3) + 1$.

\section{MATLAB Implementation and Examples}\label{sec:examples}

	The semidefinite programming method of bounding the operator norms has been implemented in MATLAB, with the script available for download from the website \url{http://www.nathanieljohnston.com/index.php/schmidtoperatornorm/}. In order to test the semidefinite programs, we will need a theoretical result to compare the computed results to. To this end, we compute analytically the $k$th operator norms of Werner states. We also look at the operator norms of randomly generated states from the Bures measure.

\subsection{Werner States}\label{sec:werner}

Here we compute the $k$th operator norms of Werner states.
Specifically, the following result shows that if $\alpha \leq 0$ then each norm coincides with $\big\|\rho_\alpha\big\|$. If $\alpha > 0$ then $\big\|\rho_\alpha\big\|_{S(1)}$ is smaller, but the rest of the norms are all equal to $\big\|\rho_\alpha\big\|$.
	\begin{prop}\label{prop:WernerSchmidt}
		Let $\rho_\alpha \in \cl{L}(\cl{H}_n) \otimes \cl{L}(\cl{H}_n)$ be a Werner state. Then
		\begin{align*}
			\big\|\rho_\alpha\big\|_{S(1)} = \frac{1 + | \min\{\alpha,0\}|}{n(n - \alpha)} \ \text{ and } \ \big\|\rho_\alpha\big\|_{S(k)} = \frac{1 + |\alpha|}{n(n - \alpha)} \quad \text{ for } \, 2 \leq k \leq n.
		\end{align*}
	\end{prop}
	\begin{proof}
		Throughout the proof, we will work with the operator $X_\alpha := n(n - \alpha)\rho_\alpha = I - \alpha S$ to simplify the algebra. To see the result when $\alpha \leq 0$, note that for any $k$,
		\[
			\big\|X_\alpha\big\|_{S(k)} = \big\|I - \alpha S\big\|_{S(k)} \leq \big\|I\big\|_{S(k)} - \alpha \big\|S\big\|_{S(k)} = 1 - \alpha,
		\]
		
		\noindent where the inequality comes from the triangle inequality and the rightmost equality comes from the fact that $\big\|S\big\|_{S(k)} = 1$, which is easily verified. To see the other inequality, choose $\ket{v} := \ket{0} \otimes \ket{0}$ and observe that
		\[
			\bra{v}X\ket{v} = (\bra{0}\otimes\bra{0})(I - \alpha S)(\ket{0}\otimes\ket{0}) = 1 - \alpha\sum_{i,j=0}^{n-1}\braket{0}{i}\braket{j}{0}\braket{0}{j}\braket{i}{0} = 1 - \alpha.
		\]
		
		On the other hand, if $\alpha \geq 0$, then for any vector $\ket{v} = \ket{a} \otimes \ket{b}$, it follows that
		\[
			\bra{v}X_\alpha\ket{v} = (\bra{a}\otimes\bra{b})(I - \alpha S)(\ket{a}\otimes\ket{b}) = 1 - \alpha (\bra{a}\otimes\bra{b})(\ket{b}\otimes\ket{a}) = 1 - \alpha |\braket{a}{b}|^2 \leq 1.
		\]
		
		\noindent Furthermore, equality can easily be seen to be attained when $\ket{v} = \ket{0} \otimes \ket{1}$, which shows that $\big\| X_\alpha \big\|_{S(1)} = 1$. To see the result for $k \geq 2$ and $\alpha \geq 0$, use the triangle inequality again to see that $\big\| X_\alpha \big\|_{S(k)} \leq 1 + \alpha$. To show that equality is attained, let $\ket{v} = \frac{1}{\sqrt{2}}(\ket{0} \otimes \ket{1} - \ket{1} \otimes \ket{0})$ and observe that $\bra{v}X_\alpha \ket{v} = 1 + \alpha$. Since $\ket{v}$ has $SR(\ket{v}) = 2$, the result follows.
	\end{proof}
	
	The performance of the semidefinite programs for the $1$-norm is analyzed in Table~\ref{table:werner}. If the transpose map is used, then we know that the semidefinite program must give exactly $\big\|\rho_\alpha\big\|_{S(1)}$ when $n = 2$, which it indeed does. In fact, the map $\Phi_1$ defined by $\Phi_1(X) = \Tr(X)I - X$ that is used as the basis of the reduction criterion~\cite{HH98} also gives the correct answer in this case. For $n = 3$, the transpose map still happens to give the correct answer, though the reduction criterion map gives a strict upper bound when $\alpha > 0$.
	
\begin{table}[h]
	\begin{center}
  \begin{tabular}{ c | c | c | c | c }
  	\hline
  	 & & & \multicolumn{2}{|c}{Upper bound computed using...} \\
  	\hline
    $n$ & $\alpha$ & Exact $\big\|\rho_\alpha\big\|_{S(1)}$ & $\Phi_1(X) = X^T$ & $\Phi_1(X) = \Tr(X)I - X$ \\
    \hline\hline
    $2$ & $1/2$ & $1/3$ & $0.3333$ & $0.3333$ \\ \hline
    $2$ & $-1/2$ & $3/10$ & $0.3000$ & $0.3000$ \\ \hline
    $3$ & $1/2$ & $2/15$ & $0.1333$ & $0.2000$ \\ \hline
    $3$ & $-1/2$ & $1/7$ & $0.1429$ & $0.1429$ \\ \hline
  \end{tabular}
	\end{center}
\caption{The exact $1$-operator norm of various Werner states as well as the computed upper bounds obtained by using the semidefinite program defined by one of two different positive linear maps.}\label{table:werner}
\end{table}

\subsection{Randomly Generated States Via The Bures Measure}\label{sec:bures}

	As another example, we consider random density operators distributed according to the Bures measure \cite{B69,U76}, which can be generated quickly via the method of~\cite{OSZ09}. We then investigate the general behaviour of the $k$th operator norms of a density operator in $\cl{L}(\cl{H}_2) \otimes \cl{L}(\cl{H}_2)$ and $\cl{L}(\cl{H}_3) \otimes \cl{L}(\cl{H}_3)$ relative to its eigenvalues.
	
\begin{figure}[h]
\begin{center}
\includegraphics[width=0.9\textwidth]{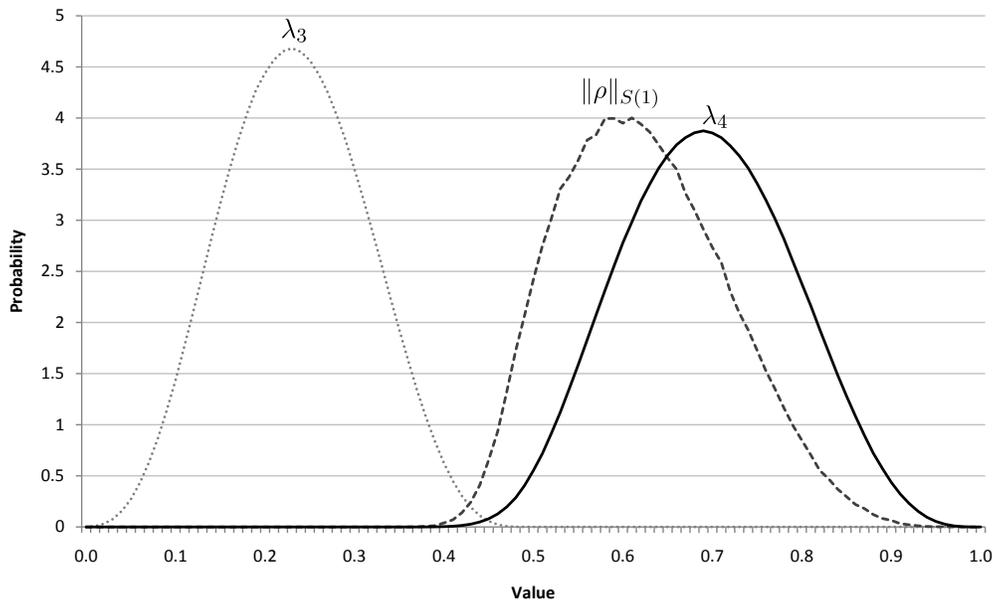}
\end{center}
\caption{Approximate distributions of the $1$-norm and the two largest eigenvalues of random Bures density operators in $\cl{L}(\cl{H}_2) \otimes \cl{L}(\cl{H}_2)$.}\label{fig:4dim}
\end{figure}

	In particular, Figure~\ref{fig:4dim} shows how the $1$-norm is distributed compared to the two largest eigenvalues $\lambda_3 \leq \lambda_4$ in $\cl{L}(\cl{H}_2) \otimes \cl{L}(\cl{H}_2)$, based on $2 \times 10^6$ randomly-generated density operators. It is not surprising that the $1$-norm lies between the $\lambda_3$ and $\lambda_4$, since $\lambda_4$ is equal to the $2$-norm, and it was shown in~\cite{JK09} that the  $(n-1)$-norm in $\cl{L}(\cl{H}_n) \otimes \cl{L}(\cl{H}_n)$ is always at least as big as the second-largest eigenvalue. We see that the $1$-norm typically is much closer to $\lambda_4$ than $\lambda_3$.

	The $1$-norm in this case was computed using the semidefinite programming method of Section~\ref{sec:semidefProgramMNorm}. A similar plot was presented in \cite{GPMSCZ09} for what was called the ``maximum local eigenvalue'', which coincides with the $1$-norm for positive operators. There it was similarly observed that the $1$-norm typically lies closer to $\lambda_4$ than $\lambda_3$ under the Hilbert-Schmidt measure.
	
\begin{figure}[h]
\begin{center}
\includegraphics[width=0.9\textwidth]{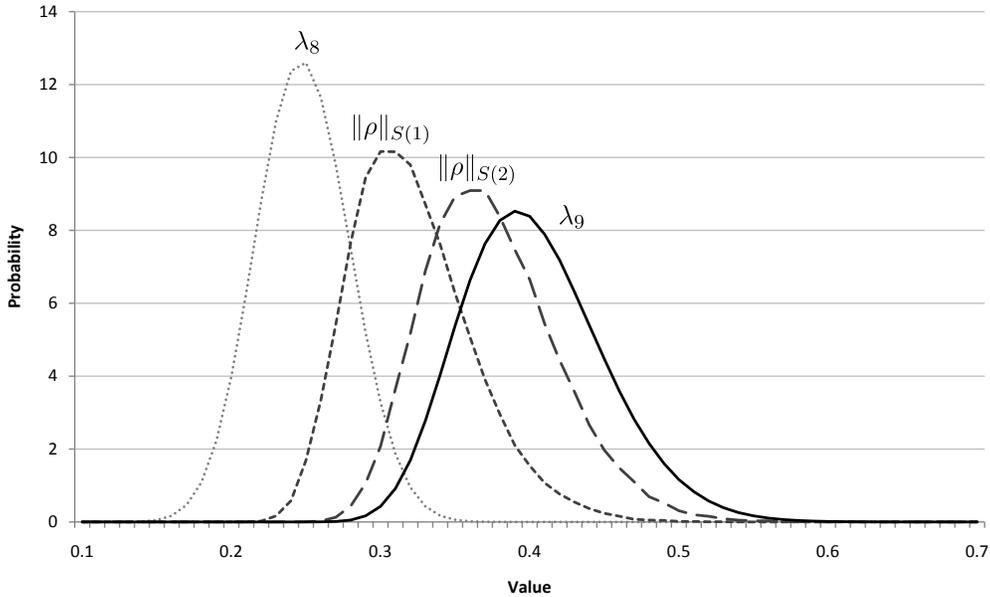}
\end{center}
\caption{Approximate distributions of the $1$ and $2$-norms, as well as the two largest eigenvalues of random Bures density operators in $\cl{L}(\cl{H}_3) \otimes \cl{L}(\cl{H}_3)$.}\label{fig:9dim}
\end{figure}

	Figure~\ref{fig:9dim} shows how the $1$ and $2$-norms typically compare to the two largest eigenvalues $\lambda_8 \leq \lambda_9$ in $\cl{L}(\cl{H}_3) \otimes \cl{L}(\cl{H}_3)$, based on $10^5$ randomly-generated density operators. As before, it is not surprising that the $2$-norm lies between $\lambda_8$ and $\lambda_9$. However, it was shown in \cite{JK09} that there exist density operators $\rho \in \cl{L}(\cl{H}_3) \otimes \cl{L}(\cl{H}_3)$ for which $\lambda_5 \leq \|\rho\|_{S(1)} < \lambda_6$. Nonetheless, this situation seems to be extremely rare, as $\|\rho\|_{S(1)}$ generally lies between $\lambda_8$ and $\lambda_9$.

Because the semidefinite programming method of Section~\ref{sec:semidefProgramMNorm} does not produce the exact value for the $1$ and $2$-norms in $\cl{L}(\cl{H}_3) \otimes \cl{L}(\cl{H}_3)$, the values of the norms used for Figure~\ref{fig:9dim} are estimates that were derived from a simple genetic algorithm.

\section{Norms Restricted to Other Convex Cones of Operators}\label{sec:coneNorms}

We will now see that many of the results for the $k$th operator norms actually hold in the much more general setting of arbitrary convex mapping cones of operators. We will begin by defining the notion of a mapping cone, which was originally introduced by St\"{o}rmer\cite{S86}.
\begin{defn}
	Let $\mathcal{S} \subseteq \mathcal{L}(\cl{L}(\cl{H}_n),\cl{L}(\cl{H}_m))$ be a cone of completely positive linear maps. $\cl{S}$ is said to be a {\em mapping cone} if $\Phi \circ \Psi \in \mathcal{S}$ and $\Psi \circ \Phi \in \mathcal{S}$ whenever $\Phi \in \mathcal{S}$ and $\Psi$ is completely positive.
\end{defn}

Mapping cones appeared recently in \cite{SSZ09} as a way of generalizing the dual cone relationships between $k$-block positive operators and operators with Schmidt number no greater than $k$. These dual relationships can be seen implicitly in the semidefinite programming results of Section~\ref{sec:semidefProgramMNorm}, so it is no surprise that the notion of mapping cones provides a natural generalization in this setting as well. Mapping cones can be defined without the restriction that they be a subset of the completely positive maps, though the definition provided will be better-suited to our purposes.

We will say that a cone of operators $\cl{C} \subseteq (\cl{L}(\cl{H}_n) \otimes \cl{L}(\cl{H}_m))^{+}$ is a mapping cone if the cone of associated linear maps (via the Choi-Jamiolkowski isomorphism) is a mapping cone. If necessary, we will specify whether we mean a mapping cone of operators or a mapping cone of linear maps, but our meaning should be clear from context.

\begin{defn}\label{defn:coneNorm}
  Let $X \in (\cl{L}(\cl{H}))^{+}$ be positive and let $\cl{C} \subseteq (\cl{L}(\cl{H}_n) \otimes \cl{L}(\cl{H}_m))^{+}$ be a closed convex cone. Then we define the \emph{$\cl{C}$-operator norm} of $X$, denoted $\big\| X \big\|_{\cl{C}}$, by
  \begin{align*}
  	\big\|X\big\|_{\cl{C}} & := \sup_{\rho \in \cl{C}} \Big\{ \Tr(X\rho) \Big\}.
  \end{align*}
\end{defn}

It is easy to see that this defines a valid norm if $\cl{C}$ is a mapping cone. It is also a norm for many other convex cones of interest -- all that needs to be checked is that $\cl{C}$ contains a full set of $n^2m^2$ linearly independent operators. Observe also that if $\cl{C} = \cl{S}_k$ and $X \geq 0$ then this definition reduces to exactly $\big\|X\big\|_{S(k)}$. The norm $\big\|X\big\|_\cl{C}$ has a similar interpretation to that of the $k$th operator norms as well. We can think of $\big\|X\big\|_\cl{C}$ as roughly measuring how close $X$ is to an operator in $\cl{C}$.

It is trivial to see that if $\cl{C} \subseteq \cl{D}$, where $\cl{D}$ is another closed convex cone, then $\big\|X\big\|_{\cl{C}} \leq \big\|X\big\|_{\cl{D}}$. In particular this implies that $\big\|X\big\|_{\cl{C}} \leq \big\|X\big\|$ always. Additionally, several of the characterizations of the $k$th operator norms carry over in an obvious way to this more general setting.

\begin{prop}\label{prop:kPosInf1_gen}
  Let $X \in (\cl{L}(\cl{H}_n) \otimes \cl{L}(\cl{H}_m))^+$ be positive. Then $cI - X \in \cl{C}^{O}$ if and only if $c \geq \big\|X\big\|_{\cl{C}}$.
\end{prop}
\begin{proof}
	By definition, $cI - X \in \cl{C}^{O}$ if and only if
	\[
		\Tr((cI - X)\rho) = c - \Tr(X\rho) \geq 0 \quad \forall \, \rho \in \cl{C}.
	\]
	
	\noindent This if true if and only if $c \geq \big\|X\big\|_{\cl{C}}$, completing the proof.
\end{proof}

Now let $X \in (\cl{L}(\cl{H}_n) \otimes \cl{L}(\cl{H}_m))^{+}$ be positive and consider the following semidefinite program.
\begin{align}\label{sp:coneNorm}
\begin{matrix}
\begin{tabular}{r l c r l}
\multicolumn{2}{c}{{\bf Primal problem}} & \quad \quad \quad & \multicolumn{2}{c}{{\bf Dual problem}} \\
\text{maximize:} & $\Tr(X\rho)$ & \quad & \text{minimize:} & $\lambda$ \\
\text{subject to:} & $\Tr(\rho) = 1$ & \quad & \text{subject to:} & $\lambda I_n \geq Y + X$ \\
\ & $\rho \in \cl{C}$ & \ & \ & $Y \in \cl{C}^{O}$ \\
\end{tabular}
\end{matrix}
\end{align}

It is easy to see that these problems are indeed duals of each other and form a valid semidefinite program, using the same method as was used in Section~\ref{sec:semidefProgramMNorm} to show that the semidefinite program~\eqref{sp:matrixNorm} is valid. Strong dual duality also holds in this setting. The main difference here is that we have $\rho \in \cl{C}$ and $Y \in \cl{C}^{O}$ rather than $\rho, Y \geq 0$ -- we could have stated the semidefinite program~\eqref{sp:matrixNorm} in terms of the cone $\cl{S}_k$, but then it would become less clear how to actually implement the semidefinite programs and compute upper bounds of $\big\|X\big\|_{S(k)}$ using $k$-positive maps.

Just as in the case for the $k$th operator norms, the theory of semidefinite programming leads to the following two results. We state them without proof, as their proofs are almost identical to the proofs of Theorem~\ref{thm:kPosInf2} and Corollary~\ref{cor:SPcor}, respectively.
\begin{thm}\label{thm:kPosInf2_gen}
	Let $X \in (\cl{L}(\cl{H}_n) \otimes \cl{L}(\cl{H}_m))^{+}$ be positive. Then
	\begin{align*}
	  \big\| X \big\|_{\cl{C}} = \inf_{Y} \big\{ \big\|X + Y\big\| : Y \in \cl{C}^{O} \big\}.
	\end{align*}
\end{thm}

\begin{cor}
	Let $Y = Y^* \in \cl{L}(\cl{H}_n) \otimes \cl{L}(\cl{H}_m)$. Then $Y \in \cl{C}^O$ if and only if
	\begin{align*}
	  \big\| X \big\|_{\cl{C}} \leq \big\|X + Y\big\| \quad \forall \, X \in (\cl{L}(\cl{H}_n) \otimes \cl{L}(\cl{H}_m))^{+}.
	\end{align*}
\end{cor}

\subsection{Application to PPT States}\label{sec:pptNorm}

Given any positive linear map $\Phi : \cl{L}(\cl{H}_n) \rightarrow \cl{L}(\cl{H}_m)$, there exists a natural convex cone $\cl{C}_\Phi$ associated with $\Phi$:
\[
	\cl{C}_\Phi := \Big\{ X \in (\cl{L}(\cl{H}_n) \otimes \cl{L}(\cl{H}_m))^{+} : (id_n \otimes \Phi)(X) \geq 0 \Big\}.
\]

\noindent Given any such convex cone, $\big\|\cdot\big\|_{\cl{C}_\Phi}$ is indeed a norm and we are able to compute $\big\|X\big\|_{\cl{C}_\Phi}$ to any desired accuracy via semidefinite programming, as seen in the previous section. In fact, $\big\|X\big\|_{\cl{C}_{\Phi_k}}$ is exactly what is computed by the semidefinite program~\eqref{sp:matrixNorm}. It follows that $\big\|X\big\|_{S(k)} = \inf_{\Phi_k}\{ \big\|X\big\|_{\cl{C}_{\Phi_k}} : \Phi_k \text{ is } \text{$k$-positive} \}$.

In the case of the transpose map $T : \cl{L}(\cl{H}_n) \rightarrow \cl{L}(\cl{H}_n)$, $\cl{C}_T$ is exactly the cone of unnormalized PPT states, and so the norm $\big\| \cdot \big\|_{\cl{C}_T}$ can be seen as a measure of how close a given operator is to having positive partial transpose. It is known \cite{S08} that the dual cone of the PPT states is given by
	\[
		\cl{C}_{T}^{O} = \Big\{ X = X^* \in \cl{L}(\cl{H}_n) \otimes \cl{L}(\cl{H}_m) : X = Y + Z \text{ for some } Y \geq 0, (id_n \otimes T)(Z) \geq 0 \Big\}.
	\]
	
\noindent This leads immediately to the following characterizations of $\big\| \rho \big\|_{\cl{C}_T}$ via Theorem~\ref{thm:kPosInf2_gen}.

\begin{prop}
	Let $\rho \in \cl{L}(\cl{H}_n) \otimes \cl{L}(\cl{H}_m)$ be a density operator. Then
	\begin{align*}
	  \big\| \rho \big\|_{\cl{C}_T} = \inf_{Y} \big\{ \big\|\rho + Y\big\| : (id_n \otimes T)(Y) \geq 0 \big\}.
	\end{align*}
\end{prop}

\section{Norms on General Projections and a Conjecture of Brandao}\label{sec:BrandaoConjecture}

We have seen that the $k$th operator norms of orthogonal projections have several applications within quantum information theory. One more reason for studying these norms comes from their appearance in a conjecture of Brandao \cite{BRANCONJ}, which asks whether or not there exists a $0 < \varepsilon < 1$ such that, for all $n$ and all orthogonal projections $P = P^* = P^2 \in \cl{L}(\cl{H}_n) \otimes \cl{L}(\cl{H}_n)$,
\begin{align}\label{eq:branPbound}
  \big\| P \big\|_{S(1)} \geq \sqrt{\frac{{\rm rank}(P)}{n^{2+\varepsilon}}}.
\end{align}

\noindent In order to answer this question, recall the inequalities~(6) and (7) of Theorem~4.13 of \cite{JK09}.

\begin{figure}[h]
\begin{center}
\includegraphics[width=\textwidth]{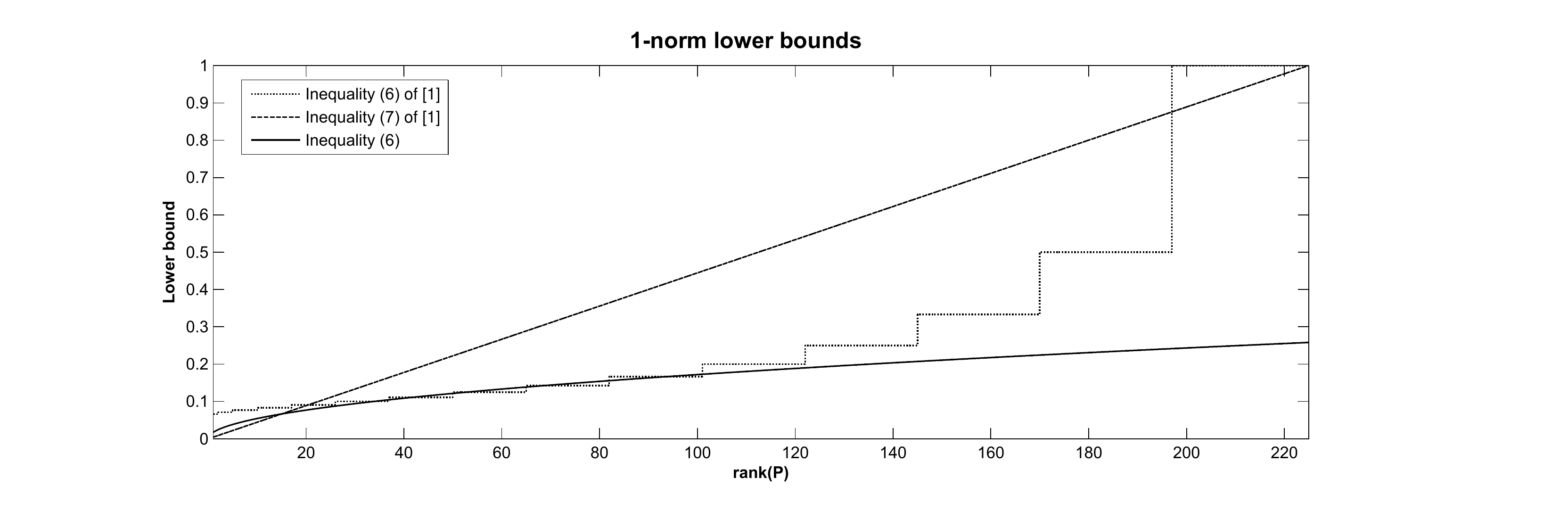}
\end{center}
\caption{A comparison of the various lower bounds for the norm on projections in the $n = m = 15$ and $k = 1$ case. Taken together, Inequalities (6) and (7) from \cite{JK09} show that Inequality~\eqref{eq:branPbound} holds for all fixed, finite $n$ (Inequality~\eqref{eq:branPbound} is shown with $\varepsilon = 0.99$).}\label{fig:proj_bound}
\end{figure}

For any \emph{fixed} $n \in \bb{N}$ and ${\rm rank}(P) \leq n$, then Inequality~(6) of \cite{JK09} implies that the statement of Inequality~\eqref{eq:branPbound} holds for some $0<\varepsilon_n < 1$. Similarly, if ${\rm rank}(P) > n$ then Inequality~(7) of \cite{JK09} implies that the statement holds for some $\varepsilon_n < 1$. Thus, the statement of Inequality~\eqref{eq:branPbound} holds in any \emph{fixed} finite dimension.

  Nonetheless, Inequality~\eqref{eq:branPbound} can be seen not to hold as $n$ tends to infinity when ${\rm rank}(P) \approx n$ via methods of convex geometry. In particular, we prove the following result using the ideas presented in \cite{ASW10}.

\begin{thm}\label{thm:genProjections}
	There exists a universal constant $C$, independent of $n$ and $k$, such that for a general orthogonal projection $P \in \cl{L}(\cl{H}_n) \otimes \cl{L}(\cl{H}_n)$ with rank$(P) \leq kn$, we have
	\begin{align*}
		\frac{k}{n} \leq \|P\|_{S(k)} \leq C\frac{k}{n}.
	\end{align*}
\end{thm}
\begin{proof}
	The left inequality is true for all projections simply by Theorem~4.13 of \cite{JK09}. We will prove the right inequality by making use of the ``tangible version'' of Dvoretzky's theorem that appears in \cite{ASW10}. First note that for any orthogonal projection $P \in \cl{L}(\cl{H}_n) \otimes \cl{L}(\cl{H}_n)$,
	\begin{align}\label{eq:Pnorm}
		\sqrt{\|P\|_{S(k)}} = \sup_{\ket{v} \in Range(P)}\Big\{ \sqrt{\sum_{i=1}^k \alpha_i^2} : \{ \alpha_i \} \text{ are the Schmidt coefficients of $\ket{v}$}\Big\}.
	\end{align}
	
	\noindent This characterization of $\|P\|_{S(k)}$ appeared in the proof of Theorem~4.15 of \cite{JK09} and also in \cite{PPHH07}, so we use it without proof. But now by associating $\cl{H}_n \otimes \cl{H}_n$ with $\cl{L}(\cl{H}_n)$, the quantity~\eqref{eq:Pnorm} equals
	\begin{align}\label{eq:Anorm}
		\sup_{A \in R}\Big\{ \sqrt{\sum_{i=1}^k s_i^2(A)} : \|A\|_F = 1, s_1(A) \geq s_2(A) \geq \cdots \geq s_n(A) \geq 0 \text{ singular values of $A$} \Big\},
	\end{align}
	
	\noindent where $R$ is the subspace of $\cl{L}(\cl{H}_n)$ associated with the range of $P$ through the standard bipartite vector to operator isomorphism. So now the goal is to show that there exists a constant $C$ such that $\sqrt{\sum_{i=1}^k s_i^2(A)} \leq C\sqrt{k/n}\|A\|_F$ for $A$ in general subspaces $R$ of dimension $kn$. To this end, we need to bound the constants $b$ and $M$ of Dvoretzky's theorem. It is trivial to see that $\sqrt{\sum_{i=1}^k s_i^2(A)} \leq \|A\|_F$ and that equality if attained for some operators $A$, so $b = 1$.
	
	To upper-bound $M$, recall from \cite{ASW10} that the expectation of the operator norm, $\bb{E}\|A\|$, is upper-bounded by $\frac{C_0}{\sqrt{n}}$ for some absolute constant $C_0$. Thus
	\begin{align*}
		M := \bb{E}\left(\sqrt{\sum_{i=1}^k s_i^2(A)}\right) \leq \bb{E}(\sqrt{k}s_1(A)) = \sqrt{k}\bb{E}\|A\| \leq C_0 \sqrt{\frac{k}{n}}.
	\end{align*}
	
	It follows via Dvoretzky's theorem that there is a constant $c$ such that if we choose $\epsilon = 1/(C_0 \sqrt{c})$, then for general subspaces $R$ with $\dim(R) \leq c \epsilon^2 C_0^2kn = kn$, we have
	\begin{align*}
		\sqrt{\sum_{i=1}^k s_i^2(A)} \leq (1 + \epsilon)M\|A\|_F \leq (1 + \frac{1}{C_0 \sqrt{c}})C_0\sqrt{\frac{k}{n}}\|A\|_F.
	\end{align*}
\end{proof}

In the case when $k = 1$, Theorem~\ref{thm:genProjections} tells us that for general projections of rank $n$, $\|P\|_{S(1)} \leq C/n$, so Inequality~\eqref{eq:branPbound} can not hold as $n$ tends to infinity if $\varepsilon < 1$. However, Inequality~\eqref{eq:branPbound} is still relevant as it only needs to hold for certain projections to have important implications. If it did, it would imply that the regularized relative entropy of entanglement \cite{VP98,VPRN97} is super-additive \cite{BRANCOMM}. This in turn would imply that $QMA(k) = QMA(2)$ for all $k > 2$ via a result of Aaronson et. al. \cite{ABDFS09}.

\vspace{0.1in}

\noindent{\bf Acknowledgements.} We thank Guillaume Aubrun, Fernando Brandao, Sevag Gharibian and Stanislaw Szarek for helpful conversations. N.J. was supported by an NSERC Canada Graduate Scholarship and the University of Guelph Brock Scholarship. D.W.K. was supported by NSERC Discovery Grant 400160, NSERC Discovery Accelerator Supplement 400233, and Ontario Early Researcher Award 048142.


\section*{Appendix I: Implementing Semidefinite Programs}\label{sec:sdpExplain}

The presentation of a semidefinite program in Section~\ref{sec:SP} was as an optimization problem defined by a Hermicity-preserving linear map $\Phi : \cl{L}(\cl{H}_n) \rightarrow \cl{L}(\cl{H}_m)$, two operators $A \in \cl{L}(\cl{H}_n)$ and $B \in \cl{L}(\cl{H}_m)$, and a closed convex cone $\cl{C} \subseteq \cl{L}(\cl{H}_n)$ with the following primal and dual forms:
\begin{align}\label{sp:formB}
\begin{matrix}
\begin{tabular}{r l c r l}
\multicolumn{2}{c}{{\bf Primal problem}} & \quad \quad \quad & \multicolumn{2}{c}{{\bf Dual problem}} \\
\text{maximize:} & $\Tr(AX)$ & \quad & \text{minimize:} & $\Tr(BY)$ \\
\text{subject to:} & $\Phi(X) \leq B$ & \quad & \text{subject to:} & $\Phi^\dagger(Y) \geq A$ \\
\ & $X \in \cl{C}$ & \ & \ & $Y \in \cl{C}^O$ \\
\end{tabular}
\end{matrix}
\end{align}

\noindent Here we will show explicitly, in the special case of $\cl{C} = (\cl{L}(\cl{H}_n))^{+}$, how to convert the above semidefinite program into the so-called standard form of a semidefinite program, defined by a vector $c \in \cl{H}_\ell$ and operators $D \in \cl{L}(\cl{H}_p)$ and $\big\{F_i\big\} \in \cl{L}(\cl{H}_p)$, with the following primal and dual forms:
\begin{align}\label{sp:formC}
\begin{matrix}
\begin{tabular}{r l c r l}
\multicolumn{2}{c}{{\bf Primal problem}} & \quad \quad \quad & \multicolumn{2}{c}{{\bf Dual problem}} \\
\text{maximize:} & $x^*c$ & \quad & \text{minimize:} & $\Tr(DY)$ \\
\text{subject to:} & $\sum_{i=1}^\ell x_iF_i \leq D$ & \quad & \text{subject to:} & $\Tr(F_iY) = c_i \quad \forall \, 1 \leq i \leq \ell$ \\
\ & \ & \ & \ & $Y \geq 0$ \\
\end{tabular}
\end{matrix}
\end{align}

Once the conversion from form~\eqref{sp:formB} to form~\eqref{sp:formC} has been carried out, the problem can be given to a semidefinite program solver to be solved. In particular, we provide a MATLAB front-end that carries out the upcoming conversion and uses the SeDuMi semidefinite program solver \cite{SeDuMi} to compute the solution.

Thus, assume that you have a semidefinite program in the form~\eqref{sp:formB}. Define a linear map $\Psi : \cl{L}(\cl{H}_n) \rightarrow (\cl{L}(\cl{H}_m) \oplus \cl{L}(\cl{H}_n))$ by
\begin{align*}
	\Psi(X) := \begin{bmatrix}\Phi(X) & 0 \\ 0 & -X\end{bmatrix}.
\end{align*}

\noindent Then the requirements that $\Phi(X) \leq B$ and $X \geq 0$ are equivalent to the single constraint
\[
	\Psi(X) \leq \begin{bmatrix}B & 0 \\ 0 & 0 \end{bmatrix}.
\]

\noindent The dual map $\Psi^\dagger : (\cl{L}(\cl{H}_m) \oplus \cl{L}(\cl{H}_n)) \rightarrow \cl{L}(\cl{H}_n)$ acts on block diagonal operators as
\begin{align*}
	\Psi^\dagger\Big(\begin{bmatrix}Y & 0 \\ 0 & Z\end{bmatrix}\Big) = \Phi^\dagger(Y) - Z.
\end{align*}

\noindent Thus, the semidefinite program~\eqref{sp:formB} can be written in the following form:
\begin{align}\label{sp:formD}
\begin{matrix}
\begin{tabular}{r l c r l}
\multicolumn{2}{c}{{\bf Primal problem}} & \quad \quad \quad & \multicolumn{2}{c}{{\bf Dual problem}} \\
\text{maximize:} & $\Tr(AX)$ & \quad & \text{minimize:} & $\Tr(DW)$ \\
\text{subject to:} & $\Psi(X) \leq D$ & \quad & \text{subject to:} & $\Psi^\dagger(W) = A$ \\
\ & \ & \ & \ & $W \geq 0$ \\
\end{tabular}
\end{matrix}
\end{align}

\noindent where $D := \begin{bmatrix}B & 0 \\ 0 & 0\end{bmatrix}$ and $W := \begin{bmatrix}Y & 0 \\ 0 & Z\end{bmatrix}$. Note in particular that we can replace the inequality in the dual problem~\eqref{sp:formB} by equality in~\eqref{sp:formD} because of the flexibility that was introduced by the arbitrary positive operator $Z$. Now let ${E_a}$ and ${F_a}$ be families of left and right generalized Choi-Kraus operators for $\Psi$ (that is, operators such that $\Psi(X) = \sum_a E_a X F_a$). Denote the $(g,h)$-entry of $X$ by $x_{gh}$ and the $(i,j)$-entry of $E_a$ and $F_a$ by $e_{aij}$ and $f_{aij}$, respectively. Then
\begin{align*}
	\Psi(X) = \sum_a E_a X F_a = \sum_a\Big( \sum_{gh}e_{aig}x_{gh}f_{ahj} \Big)_{ij} = \sum_{gh}x_{gh}G_{gh},
\end{align*}

\noindent where
\begin{align*}
	G_{gh} := \sum_a\begin{bmatrix}e_{a1g}f_{ah1} & e_{a1g}f_{ah2} & \cdots & e_{a1g}f_{ah(m+n)} \\
	e_{a2g}f_{ah1} & e_{a2g}f_{ah2} & \cdots & e_{a2g}f_{ah(m+n)} \\
	\vdots & \vdots & \ddots & \vdots \\
	e_{a(m+n)g}f_{ah1} & e_{a(m+n)g}f_{ah2} & \cdots & e_{a(m+n)g}f_{ah(m+n)}\end{bmatrix}.
\end{align*}

Then by examining the equality constraint in the SDP~\eqref{sp:formD}, we see that, for all $g,h$,
\begin{align*}
	a_{gh} & = \big(\Psi^\dagger(W)\big)_{gh} = \Big( \sum_{a}F_a W E_a \Big)_{gh} = \sum_a\sum_{ij}f_{agi}w_{ij}e_{ajh} \\
	& = \sum_a \Tr\Big( \begin{bmatrix} \sum_i f_{agi}w_{i1}e_{a1h} & \sum_i f_{agi}w_{i2}e_{a1h} & \cdots & \sum_i f_{agi}w_{i(m+n)}e_{a1h} \\ \sum_i f_{agi}w_{i1}e_{a2h} & \sum_i f_{agi}w_{i2}e_{a2h} & \cdots & \sum_i f_{agi}w_{i(m+n)}e_{a2h} \\ \vdots & \vdots & \ddots & \vdots \\ \sum_i f_{agi}w_{i1}e_{a(m+n)h} & \sum_i f_{agi}w_{i2}e_{a(m+n)h} & \cdots & \sum_i f_{agi}w_{i(m+n)}e_{a(m+n)h} \end{bmatrix} \Big) \\
	& = \Tr(G_{hg}W).
\end{align*}

It follows that the semidefinite program~\eqref{sp:formD} can be written in the form:
\begin{align}\label{sp:formE}
\begin{matrix}
\begin{tabular}{r l c r l}
\multicolumn{2}{c}{{\bf Primal problem}} & \quad \quad \quad & \multicolumn{2}{c}{{\bf Dual problem}} \\
\text{maximize:} & ${\rm vec}(X)^*{\rm vec}(A)$ & \quad & \text{minimize:} & $\Tr(DW)$ \\
\text{subject to:} & $\sum_{gh}^\ell x_{hg}G_{hg} \leq D$ & \quad & \text{subject to:} & $\Tr(G_{hg}W) = a_{gh} \quad \forall \, 1 \leq g,h \leq n$ \\
\ & \ & \ & \ & $W \geq 0$ \\
\end{tabular}
\end{matrix}
\end{align}

\noindent where ${\rm vec}(X)$ and ${\rm vec}(A)$ are the vectorizations of $X$ and $A$, respectively, that are obtained by stacking the columns of the matrices on top of each other into a column vector in the usual way. The semidefinite program~\eqref{sp:formE} is in standard form, so it can now be input into a semidefinite program solver.

This transformation of a semidefinite program of the form~\eqref{sp:form} into a semidefinite program in standard form has been implemented in MATLAB as a front-end for the SDP solver SeDuMi. The code and usage instructions can be downloaded from \url{http://www.nathanieljohnston.com/index.php/quantumsedumi/}.
\end{document}